\documentclass[
reprint,
a4paper,
australian,
amsmath,
amssymb,
aps,
prd,
twocolumn,
superscriptaddress,
preprintnumbers,
nofootinbib
]{revtex4-2}

\usepackage{footnote} 
\usepackage{braket}
\usepackage{cancel}
\usepackage{lipsum}
\usepackage{tikz, pgf}
\usetikzlibrary{positioning}
\usepackage{graphicx}
\usepackage{amssymb}
\usepackage{amsmath}
\usepackage{color,xcolor}
\definecolor{darkgreen}{rgb}{0.1,0.4,0.0}
\usepackage{times}
\usepackage{mathrsfs}
\usepackage[utf8]{inputenc}
\usepackage[normalem]{ulem}
\usepackage{feynmp-auto}
\usepackage{subcaption}
\usepackage{tabularx}
\usepackage{float}
\usepackage{booktabs} 
\usepackage{hyperref} 
\usepackage[capitalise]{cleveref}
\usepackage{multirow} 

\newcommand{\ome}{\omega}

\newcommand{\Lam}{\Lambda}

\newcommand{\eps}{\epsilon}
\newcommand{\rr}{\to}
\newcommand{\ph}{\gamma}
\newcommand{\dph}{\gamma\gamma}

\newcommand{\f}[2]{\frac{#1}{#2}}

\newcommand{\sref}[1]{Sec.~\ref{#1}}
\newcommand{\splittwo}[3]{
    \begin{align}
    &#1 \nonumber \\
    &\times #2 
    \label{#3}
    \end{align}
}
\newcommand{\splitthree}[3]{
    \begin{align}
    &#1 \nonumber \\
    &\times #2 \nonumber \\
    &\times #3 
    \end{align}
}

\hypersetup{
	colorlinks=true,       
	linkcolor=blue,        
	citecolor=blue,        
	filecolor=magenta,     
	urlcolor=blue         
}

\begin{document}
\title{Two-photon production of \texorpdfstring{$f_0$}{f0} and \texorpdfstring{$a_0$}{a0} resonances as hadronic molecules composed of two vector mesons}
\author{Li-Ke Yang}
\email{yanglike@itp.ac.cn}
\affiliation{CAS Key Laboratory of Theoretical Physics, Institute of Theoretical Physics, Chinese Academy of Sciences, Beijing 100190, China}
\affiliation{School of Physics, University of Chinese Academy of Sciences (UCAS), Beijing 100049, China}

\author{Zheng-Li Wang}
\email{wangzhengli@ucas.ac.cn}
\affiliation{School of Physics, University of Chinese Academy of Sciences (UCAS), Beijing 100049, China}

\author{Jia-Jun Wu}
\email{wujj@ucas.ac.cn}
\affiliation{School of Physics, University of Chinese Academy of Sciences (UCAS), Beijing 100049, China}

\author{Bing-Song Zou}
\email{zoubs@itp.ac.cn}
\affiliation{Department of Physics, Tsinghua University, Beijing 100084, China}
\affiliation{CAS Key Laboratory of Theoretical Physics, Institute of Theoretical Physics, Chinese Academy of Sciences, Beijing 100190, China}
\affiliation{School of Physics, University of Chinese Academy of Sciences (UCAS), Beijing 100049, China}
\begin{abstract}
Ascribed as $\rho\rho$ and $K^* \bar K^*$ molecular states, respectively, iso-scalar $f_0(1500)$ and $f_0(1710)$ states are expected to have iso-vector partners, potentially identified as $a_0(1450)$ and $a_0(1710)$. The predicted dominant decay modes for these two $a_0$ resonances are $a_0(1450)\to\omega\pi\pi$ and $a_0(1710)\to\omega\pi\pi,\,\phi\pi\pi$. We estimate cross sections for two-photon production of these four resonances within hadronic molecular picture, and demonstrate that SuperKEKB's luminosity is sufficient for their observation and more precise parameter measurements. 
\end{abstract}
\maketitle
\section{Introduction}
Understanding the hadron spectrum is one of the most important challenges in hadron physics. There are various models interpreting the observed states, and selecting the right one is a key issue. These models can be differentiated by examining their predictions of the complete spectrum, with consideration of fundamental symmetries. For example, in the physical depiction of $K\bar K$ molecules, both iso-scalar and iso-vector states are symmetrically generated, corresponding to the existing states $f_0(980)$ and $a_0(980)$~\cite{Oller:1997ti, Guo:2017jvc}.

The $f_0(1370)$ was previously considered to be an iso-scalar $\rho\rho$ molecular state, dynamically generated by the $\rho\rho$ interaction~\cite{Geng:2008gx,Molina:2008jw}. However, the inclusion of coupled-channels of pseudoscalar mesons suggests that the pole and partial decay widths are more closely aligned with those of the $f_0(1500)$ listed in the Review of Particle Physics (RPP)~\cite{Wang:2019niy}.

Similarly, $f_0(1710)$ is interpreted as the iso-scalar $K^*\bar K^*$ molecular state, dynamically generated by the vector meson-vector meson interaction~\cite{Geng:2008gx,Du:2018gyn}.
Including the coupled-channels of pseudoscalar mesons does not significantly shift the pole, and the mass, width, and partial widths are consistent with those of $f_0(1710)$ reported in the RPP~\cite{Wang:2021jub}.

However, the measurement as well as exploration of the iso-vector partners of $f_0(1500)$ and $f_0(1710)$ as hadronic molecules are less explicit. Given that $\rho\rho$ cannot form an $a_0$ state within the isospin formalism due to Bose-Einstein statistics, which requires $l+S+I=\text{even}$, $\rho\omega$ emerges as the most promising candidate for forming the iso-vector partner of $f_0(1500)$. This partner is postulated to be the $a_0(1450)$~\cite{Wang:2022pin}. Meanwhile, the iso-vector partner of $f_0(1710)$ is still missing, potentially identifiable as $a_0(1710)$~\cite{Wang:2022pin}.

Therefore, to validate the vector meson-vector meson molecule model of $f_0(1500)$ and $f_0(1710)$ , it is essential to confirm the nature of both $a_0(1450)$ and $a_0(1710)$. However, up to now, different laboratories have not reached a consensus on the measured masses of $a_0(1450)$ and $a_0(1710)$. The situation is even worse when it comes to the measurement of the width. For $a_0(1450)$, measurements vary across different experiments, while for $a_0(1710)$, only two results have been reported so far.

A notable measurement of $a_0(1450)$ is from Belle in the study of the reaction $\gamma\gamma \to \eta\pi^0$~\cite{Belle:2009xpa}. The partial-wave analysis of the differential cross section indicates the presence of a resonance that could be possibly attributed to $a_0(1450)$. However, its nominal fit mass is $1316.8\,\text{MeV}$, with no corresponding peak for $a_0(1450)$ in the total cross section spectrum. In fact, if $a_0(1450)$ is a hadronic molecule, its dominant decay channel is predicted to be $\omega\pi\pi$~\cite{Wang:2022pin}, so investigating $\gamma\gamma\to \omega\pi\pi$ is crucial to finding $a_0(1450)$.

The BaBar Collaboration observed $a_0(1700)\to \eta\pi$ in their $\dph\to\eta_c\to \eta\pi^+\pi^-$ analysis~\cite{BaBar:2021fkz}. The BESIII Collaboration also found evidence supporting the existence of an iso-vector partner of $f_0(1710)$~\cite{BESIII:2021anf}, reinforcing the existence of $a_0(1710)$ with a mass close to that of $f_0(1710)$. However, a newly observed $a_0$-like state with a mass of $1.817\,\text{GeV}$ is also suspected to be $a_0(1710)$~\cite{BESIII:2022npc}.

Therefore, it is important to identify suitable reactions for the search of $a_0(1450)$ and $a_0(1710)$ to obtain more accurate measurements. The two-photon production reaction could be an ideal place to detect $a_0(1450)$, $a_0(1710)$, as well as $f_0(1500)$ and $f_0(1710)$. In this process, one can exclude additional hadrons in the final state, except for the decay products of the target resonance, such as observing $a_0(1450)$ in $\dph\rr \ome\pi^+\pi^-,\eta\pi^0$. Moreover, with the full upgrade of KEKB to SuperKEKB, it is expected to provide 40 times more two-photon production events compared to KEKB~\cite{Funakoshi:2023wgx}.

In this paper, we calculate the cross sections for the two-photon production of $a_0(1450)$, $a_0(1710)$, $f_0(1500)$, and $f_0(1710)$, followed by their subsequent decays into main decay channels, exemplified by $\gamma\gamma\to a_0(1450)\to \omega\pi\pi$. We also present the expected $N_p$s for these reactions at SuperKEKB. The paper is organized as follows: In \sref{sec:Formalism}, we mainly present the Lagrangians and other needed definitions. In \sref{sec:Results}, we show the numerical results and discussions. Finally we conclude with a summary in \sref{sec:Summary}.

\section{Formalism}\label{sec:Formalism}
\subsection{The Interaction Vertices}
The effective Lagrangians involved are as follows~\cite{Nagahiro:2008cv}:
\begin{align}
    \mathcal L_{VVV} &= ig\langle (\partial_\mu V_\nu - \partial_\nu V_\mu) V^\mu V^\nu \rangle, \\
    \mathcal L_{V\ph} &= - M_V^2 \frac{e}{g} A_\mu\langle V^\mu Q\rangle, \\
    \mathcal L_{VPP} &= -ig\langle V^\mu[P,\partial_\mu P]\rangle,\\
    \mathcal L_{V_1V_2a_0} &= g_{a_0V_1V_2} \phi_{a_0} V_1^\mu V_{2\mu}, \\
    \mathcal L_{V_1V_2f_0} &= g_{f_0V_1V_2} \phi_{f_0} V_1^\mu V_{2\mu}.
\end{align}
$g=M_V/2f$ is the hidden gauge coupling constant, where $f=93\,\text{MeV}$ the pion decay constant and $M_V$ is the mass of the vector meson $V$; $\langle ... \rangle$ denotes the trace in SU(3) space; $Q=diag(2,-1,-1)/3$, and $e=-|e|$ is the electron charge; $\phi_S$ is the field of the scalar resonance $S$; $V_i^\mu$, with $i=1,2$, are the fields of the vector mesons $V_i$. Additionally, $V_\mu$ is the vector field matrix defined as follows:
\begin{equation}
\left(\begin{array}{ccc} \frac{1}{\sqrt 2}\rho^0 + \frac{1}{\sqrt 2}\omega & \rho^+ & K^{*+} \\ \rho^- & -\frac{1}{\sqrt 2}\rho^0 + \frac{1}{\sqrt 2}\omega & K^{*0} \\ K^{*-} & \bar K^{*0} & \phi  \end{array}\right)_\mu,
\end{equation}
and $P$ is the pseudoscalar field matrix defined as follows:
\begin{equation}
    \left(\begin{array}{ccc}
    \frac{1}{\sqrt2}\pi^0+\frac{1}{\sqrt6}\eta &
    \pi^+ & K^{+} \\
    \pi^- & -\frac{1}{\sqrt2}\pi^0+\frac{1}{\sqrt6}\eta & K^{0} \\
    K^- & \bar K^0 & -\frac{2}{\sqrt6}\eta
    \end{array}\right).
\end{equation}

\begin{figure}
  \centering
  \begin{subfigure}[htbp]{0.425\textwidth}
    \centering
    \includegraphics[width=\textwidth]{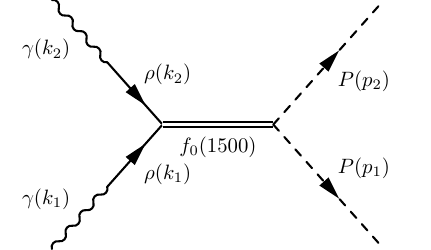}
    \caption{}
    \label{fig:feynman1}
  \end{subfigure}
  \begin{subfigure}[htbp]{0.425\textwidth}
    \centering
    \includegraphics[width=\textwidth]{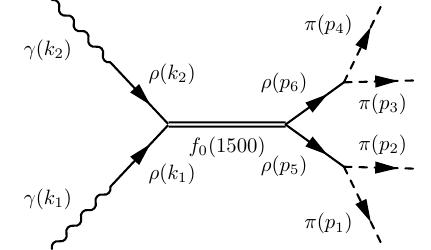}
    \caption{}
    \label{fig:feynman2}
  \end{subfigure}
  \caption{Feynman diagrams for two-photon production of $f_0(1500)$}
  \label{fig:feynman-f0(1500)}
\end{figure}

\begin{figure}
  \centering
  \begin{subfigure}[htbp]{0.425\textwidth}
    \centering
    \includegraphics[width=\textwidth]{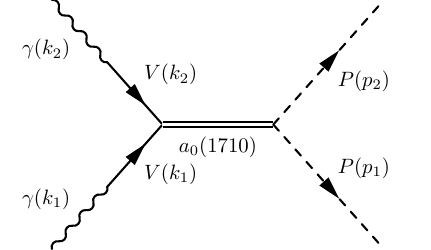}
    \caption{}
    \label{fig:feynman7}
  \end{subfigure}
  \begin{subfigure}[htbp]{0.425\textwidth}
    \centering
    \includegraphics[width=\textwidth]{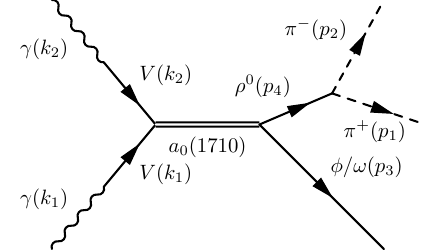}
    \caption{}
    \label{fig:feynman8}
  \end{subfigure}
  \caption{Feynman diagrams for two-photon production of $a_0(1710)$}
  \label{fig:feynman-a0(1710)}
\end{figure}

According to the Lagrangians, we can define the following tensors corresponding to different vertex structures:
\begin{align}
    (V_{SV_1V_2})^{\mu\nu} &= g_{SV_1V_2} g^{\mu\nu}, \\
    V_{SP_1P_2} &= g_{SP_1P_2}, \\
    (V_{V_1V_2V_3})^{\alpha\mu\nu} &= g_{V_1V_2V_3}\left\{(k_{V_1}^\nu-k_{V_2}^\nu)g^{\alpha\mu}+\right. \nonumber \\
    &\left. (k_{V_2}^\alpha - k_{V_3}^\alpha)g^{\mu\nu}+(k_{V_3}^\mu - k_{V_1}^\mu)g^{\nu\alpha}\right\},\\ 
    (V_{\gamma V})^{\mu\nu} &= g_{\gamma V}g^{\mu\nu}, \\
    (V_{VP_1P_2})^\mu &= g_{VP_1P_2}(k^\mu_{P_1} - k^\mu_{P_2}).
\end{align}
Note that we assume all particles in $(V_{V_1V_2V_3})^{\alpha\mu\nu}$ to be incident particles approaching the vertex. 
In practice, note that for $(V_{\rho\pi\pi})^\mu$ we have: 
\begin{align}
    (V_{\rho\pi\pi})^\mu &=(V_{\rho^0\pi^+\pi^-})^\mu + (V_{\rho^+\pi^0\pi^+})^\mu + (V_{\rho^-\pi^0\pi^-})^\mu.
\end{align}
Based on these vertices, we can construct the corresponding Feynman diagrams, as shown in Fig.~\ref{fig:feynman1}$\sim$\ref{fig:feynman8}. P refers to pseudoscalar meson and V refers to vector meson. All the detailed involved channels can be found in \sref{sec:Results}. 

\begin{figure}
  \centering
    \begin{subfigure}[htbp]{0.425\textwidth}
    \centering
    \includegraphics[width=\textwidth]{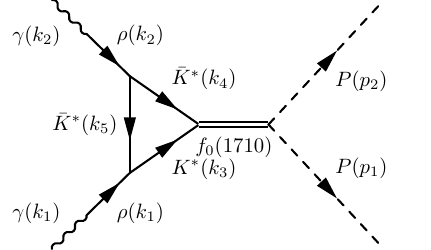}
    \caption{}
    \label{fig:feynman3}
  \end{subfigure}
    \begin{subfigure}[htbp]{0.425\textwidth}
    \centering
    \includegraphics[width=\textwidth]{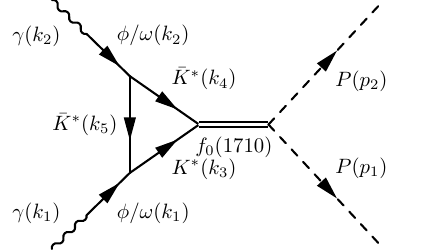}
    \caption{}
    \label{fig:feynman4}
  \end{subfigure}
  \caption{Feynman diagrams for two-photon production of $f_0(1710)$}
  \label{fig:feynman-f0(1710)}
\end{figure}

\begin{figure}
  \centering
  \begin{subfigure}[htbp]{0.425\textwidth}
    \centering
    \includegraphics[width=\textwidth]{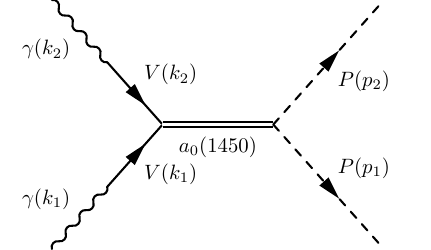}
    \caption{}
    \label{fig:feynman5}
  \end{subfigure}
  \begin{subfigure}[htbp]{0.425\textwidth}
    \centering
    \includegraphics[width=\textwidth]{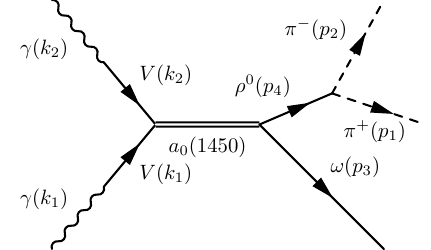}
    \caption{}
    \label{fig:feynman6}
  \end{subfigure}
  \caption{Feynman diagrams for two-photon production of $a_0(1450)$}
  \label{fig:feynman-a0(1450)}
\end{figure}

\subsection{Propogators}
The propogators for spin 0 resonance and vector meson are defined as follows:
\begin{align}
    &G_R(k) = \frac{1}{k^2-m_R^2+im_R\Gamma_R}, \\
    &(G_V(k))_{\mu\nu} = \frac{-\tilde g_{\mu\nu}}{k^2 - m_V^2+im_V\Gamma_R}, 
\end{align}
where k is the 4-momentum, and $m,\,\Gamma$ are the mass and total width of the particle. Particularly, in the case of a vector meson in $\gamma V$ vertex, where the 4-momentum satisfies $k^2=0$, the propagator is as follows:
\begin{align}
    &(G'_{V})_{\mu\nu} = \frac{ g_{\mu\nu}}{m_V^2}. 
\end{align}

Definition of $\tilde g_{\mu\nu}$ is as follows~\cite{Zou:2002ar}:
\begin{align}
\sum\limits_{m=1}^3\epsilon_\mu(m)\epsilon^{*}_\nu(m)       &= -g_{\mu\nu} + \f{p_\mu p_\nu}{p^2}\nonumber\\
     &\equiv -\tilde g_{\mu\nu}(p),
\end{align}
where $\epsilon_\mu(m)$ is the polarization vector for a massive vector meson with its 4-momentum to be $p_\mu$ and its thrid component of spin in its rest frame to be $m$.

Next, consider the photon polarization vector $A_\mu$ with 4-momentum $q_\mu$. We take into account both the Lorentz orthogonality condition $A_\mu q^\mu=0$ and an additional gauge invariance condition. We assume the Coulomb gauge $A_\mu Q^{\mu}=0$ in the c.m.~frame, where $Q^{\mu}$ is the summation of the momentum of two incoming photons. Let $K_\mu=Q_\mu-q_\mu$, the 4-momentum of the other incoming photon. The summation of polarization vectors is then given as follows~\cite{Zou:2002ar}:
\begin{align}
    \sum_{m} A_\mu(m)A^{*}_\nu(m) &= -g_{\mu\nu} + \frac{q_\mu K_\nu + K_\mu q_\nu}{q\cdot K} \nonumber\\
    &\equiv -g_{\mu\nu}^{\bot\bot}. 
\end{align}
 
\subsection{Form Factors}
For each triangle loop in the Feynman diagram as shown in Fig.~\ref{fig:feynman-f0(1710)} and Fig.~\ref{fig:feynmanLam3}, we introduce two types of form factors. The first type is a form factor with a smooth cutoff that serves to regularize the loop integration: 
\begin{equation}
    f_\Lam(k) = \exp\left[{-\f{|\vec k|^2}{(\Lam_1)^2}}\right],
\end{equation}
where $\Lambda_1=1.0\,\text{GeV}$~\cite{Wang:2021jub}. 
The second type is a dipole form factor for the off-shell exchanged meson $\bar K^*$:
\begin{equation}
    F_{K^*}(k)=\frac{(\Lambda_2)^4}{(\Lambda_2)^4+(k^2-m_{K^*}^2)^2},
\end{equation}
where $\Lambda_2=1.0\,\text{GeV}$.
Note that the empirical values for the cut-off parameters in such hadronic form factors range from 0.5 to 2.0 GeV. For simplicity, we take both $\Lambda_1$ and $\Lambda_2$ to be $1.0\,\text{GeV}$ in our calculation of the triangle hadronic loops.

For the off-shell mesons $\rho$, $\omega$, and $\phi$ in the $\gamma V$ vertex, we also introduce a dipole form factor:
\begin{equation}
    F'_V(p)=\frac{(\Lambda_3)^4}{(\Lambda_3)^4+(p^2-m_V^2)^2},
\end{equation}
where $\Lambda_3=0.63\text{--}0.83\,\text{GeV}$. The determination of $\Lambda_3$ will be discussed in \sref{subsec:cutoff para}. Besides, notice that  $p^2=0$ for the mesons coupling to the incoming photons, and we shall simply denote $F'_V(p)$ as $F'_V$.

\subsection{The Scattering Amplitudes}
We can define a function $G(k_1,k_2,k_5)$, which combines the propogators in the loop with the aforementioned form factors, to calculate the triangle loop $K^*(k_3)-\bar K^*(k_4)-\bar K^*(k_5)$, where $\bar K^*(k_5)$ is the exchanged meson, and $k_3 = k_1 + k_5$, $k_4 = k_2 - k_5$: 
 \splitthree
 {G(k_1,k_2,k_5) = F_{K^*}(k_5)\f{1}{k_5^2 - m_{\bar K^*}^2 + i\eps}}
 {\f{1}{(k_1+k_5)^2-m_{K^*}^2 + i\eps}\f{1}{(k_2-k_5)^2 - m_{\bar K^*}^2 + i\eps} }
 {\f{(\Lam_2)^4}{(\Lam_2)^4+(k_5^2-m_{\bar K^*}^2)^2} \exp\left[{-\f{|\vec k_5|^2}{(\Lam_1)^2}}\right].}
 {}
 
Next, we proceed to write down the scattering amplitude. In this section, we present one scattering amplitude as an example, while the remaining scattering amplitudes can be found in Appendix~\ref{sec:Appendix B}.

For the Feynman diagram as shown in Fig.~\ref{fig:feynman3}, the corresponding scattering amplitude of $\ph(k_1)\ph(k_2) \rr f_0(1710) \rr \pi(p_1) \pi(p_2)$ is as follows:
\begin{align}
\mathcal M_{2a} &= A_{1\mu}(m_1,k_1) A_{2\xi}(m_2,k_2)(F'_\rho)^2 \nonumber \\ 
&\times (V_{\gamma(k_1)\rho})^{\mu\nu} (V_{\gamma(k_2)\rho})^{\xi\sigma}(G'_{\rho})_{\nu\alpha} (G'_{\rho})_{\sigma\kappa} \nonumber \\ 
&\times (V_{f_0 K^* \bar K^*})^{\delta\eta} G_{f_0}(k_{1}+k_{2}) V_{f_0 P_1 P_2}\nonumber\\
&\times\int\frac{id^4 k_5}{(2\pi)^4} (-\tilde g_{\beta\delta}(k_{3}))(- \tilde g_{\lambda\eta}(k_{4}))(-\tilde g_{\gamma\theta}(k_{5})) \nonumber\\
&\times G(k_1,k_2,k_5)\left\{ (V_{\rho K^{*+}K^{*-}} )^{\alpha\beta\gamma} (V_{\rho K^{*-}K^{*+}})^{\kappa\lambda\theta}\right.\nonumber\\
&+ (V_{\rho K^{*-}K^{*+}} )^{\alpha\beta\gamma} (V_{\rho K^{*+}K^{*-}})^{\kappa\lambda\theta}\nonumber\\
&+ (V_{\rho K^{*0}\bar K^{*0}})^{\alpha\beta\gamma} (V_{\rho\bar K^{*0}K^{*0}})^{\kappa\lambda\theta}\nonumber\\
&+ \left.(V_{\rho\bar K^{*0} K^{*0}})^{\alpha\beta\gamma} (V_{\rho K^{*0}\bar K^{*0}})^{\kappa\lambda\theta}\right\}, 
\end{align}
where $A_{1\mu},A_{2\xi}$ are the polarization vectors of the two incoming photons.

\subsection{Coupling Constants}
The coupling constants $g_{\gamma V}$ involved in the tensor $(V_{\gamma V})^{\mu\nu}$ are given as follows:
\begin{align}
    g_{\gamma\rho}&=-\frac{1}{\sqrt2}M_\rho^2\frac{e}{g},\\
    g_{\gamma\omega}&=-\frac{1}{3\sqrt2}M_\ome^2\frac{e}{g},\\
    g_{\gamma\phi}&=\frac13M_\phi^2\frac{e}{g}.
\end{align}

The coupling constants $g_{V_1V_2V_3}$ involved 
in the tensor $(V_{V_1V_2V_3})^{\mu\nu\alpha}$ are given as follows:
\begin{align}
    &g_{\rho^0K^{*+}K^{*-}}=g_{\rho^0K^{*-}K^{*+}}=\frac{1}{\sqrt2}ig,\\
    &g_{\rho^0K^{*0}\bar K^{*0}}=g_{\rho^0\bar K^{*0}K^{*0}}=-\frac{1}{\sqrt2}ig,\\
    &g_{\ome K^{*+}K^{*-}}=g_{\ome K^{*-}K^{*+}}=\frac{1}{\sqrt2}ig,\\
    &g_{\ome K^{*0}\bar K^{*0}}=g_{\ome \bar K^{*0} K^{*0}}=\frac{1}{\sqrt2}ig,\\
    &g_{\phi K^{*+}K^{*-}}=g_{\phi K^{*-}K^{*+}}=ig,\\
    &g_{\phi K^{*0}\bar K^{*0}}=g_{\phi\bar K^{*0} K^{*0}}=ig.
\end{align}

The coupling constant $g_{\rho\pi\pi}$ involved in the tensor $(V_{\rho\pi\pi})^\mu$ is given as follows:
\begin{equation}
    g_{\rho\pi\pi} = -\sqrt2 ig.
\end{equation}

We now come to discuss the coupling constants involved in the tensors $(V_{SV_1V_2})^{\mu\nu}$ and $V_{SP_1P_2}$.

For a pure molecular state composed of two constituent particles, its coupling constant with the constituent particles is given by the following formula~\cite{Guo:2008zg}:
\begin{equation}
    \frac{g_{RV_1V_2}^2}{4\pi} = \frac{(m_1+m_2)^{\f52}}{(m_1m_2)^{\f12}}\sqrt{32\epsilon},
\label{eq33}
\end{equation}
where $m_1,\,m_2$ are the masses of the two component particles $V_1,\,V_2$, and $\epsilon$ is the binding energy. With this formula we can obtain $g_{a_0(1450)\rho\ome}$, $g_{a_0(1710)K^*\bar K^*}$, $g_{a_0(1710)\rho\phi}$, $g_{f_0(1500)\rho\rho}$ and $g_{f_0(1710)K^*\bar K^*}$. 
Note that $g_{f_0(1500)\rho\rho}$ gives the coupling constant of $f_0(1500)$ to both $\rho^0\rho^0$ and $\rho^+\rho^-$. Specifically, we have $g_{f_0(1500)\rho^0\rho^0} = \frac{1}{\sqrt2}g_{f_0(1500)\rho^+\rho^-}=\frac{1}{\sqrt{3}}g_{f_0(1500)\rho\rho}$. 

The other coupling constants are deduced from the relavent partial decay widths. For two-body decays, this determination can be performed using the following formula~\cite{PDG}:
\begin{equation}
    \Gamma = \f{1}{8\pi}|\mathcal M|^2 \frac{|\vec p_1|}{M^2}.
\end{equation}

For our calculations, the partial widths of $f_0(1500)$ are based on the RPP recommended values or average values~\cite{PDG}, whereas the partial widths of $f_0(1710)$ are derived using the unitary coupled channel approach, as predicted in ~\cite{Wang:2021jub}. Similarly, the partial widths for $a_0(1450)$ and $a_0(1710)$, given by the same approach, are taken from ~\cite{Wang:2021jub} and ~\cite{Wang:2022pin}, respectively. All the necessary partial
widths and the coupling constants obtained using this method are listed in Table I and Table II.

Additionally, it is worth mentioning that the mass and total width of the resonances $f_0(1500)$, $f_0(1710)$, $a_0(1450)$, and $a_0(1710)$, used for cross section calculations in \sref{sec:Results}, are presented in Table~\ref{MassAndWidth}. The data for $f_0(1500)$ and $f_0(1710)$ come from the RPP~\cite{PDG}, while the data for $a_0(1450)$ and $a_0(1710)$ are taken from ~\cite{Wang:2022pin}.

\begin{table}[htbp]
\centering
\caption{Partial widths and coupling constants relevant to $f_0(1500)$ and $f_0(1710)$}
\label{cp&decay:f0(1500)&f0(1710)}
\begin{tabular}{ccccc}
\toprule
Channels & \multicolumn{2}{c}{Partial Width [MeV]}& \multicolumn{2}{c}{Coupling Constant [GeV]} \\ 
 & $f_0(1500)$ & $f_0(1710)$ & $f_0(1500)$ & $f_0(1710)$ \\
\midrule
$\pi\pi$ & $37.3$ & $12.7$ & $1.70$ & $1.06$ \\
$K\bar K$ & $9.2$ & $44.0$ & $0.96$ & $2.16$  \\
$\eta\eta$ & $\backslash$ & $12.9$ & $\backslash$ & $1.21$ \\
\bottomrule
\end{tabular}
\end{table}

\begin{table}[htb]
\centering
\caption{Partial widths and coupling constants relevant to $a_0(1450)$ and $a_0(1710)$}
\label{cp&decay:a0(1450)&a0(1710)}
\begin{tabular}{ccccc}
\toprule  
Channels & \multicolumn{2}{c}{Partial Width [MeV]}& \multicolumn{2}{c}{Coupling Constant [GeV]} \\ 
 & $a_0(1450)$ & $a_0(1710)$ & $a_0(1450)$ & $a_0(1710)$ \\
\midrule 
$\rho\omega$ & $\backslash$ & $61.0$ & $\backslash$ & $3.50$  \\
$\eta\pi$ & $14.7$ & $66.9$ & $1.14$ & $2.55$  \\
$K\bar K$ & $13.2$ & $74.4$ & $1.15$ & $2.81$  \\
\bottomrule 
\end{tabular}
\end{table}

\begin{table}[H]
\centering
\caption{Mass and total width of $f_0$ and $a_0$ resonances}
\label{MassAndWidth}
\begin{tabular}{ccc}
\toprule
 Resonances & mass[GeV] & total width[GeV] \\
\midrule
$f_0(1500)$ & $1.522$ & $0.108$ \\
$f_0(1710)$ & $1.733$ & $0.150$ \\
$a_0(1450)$ & $1.500$ & $0.128$ \\
$a_0(1710)$ & $1.720$ & $0.220$ \\
\bottomrule
\end{tabular}
\end{table}

\subsection{Cutoff parameter \texorpdfstring{$\Lam_3$}{Lambda3}}\label{subsec:cutoff para}
\begin{figure}[htbp]
    \centering
    \includegraphics[width=0.45\textwidth]{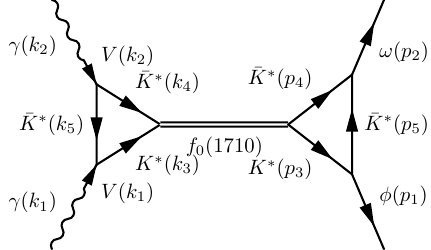}
    \caption{Feynman diagram for $\ph\ph\rr f_0(1710)\rr \phi\ome$}
    \label{fig:feynmanLam3}
\end{figure}

Belle has investigated $\ph\ph\rr VV$ and presented the cross section spectrum for $\ph\ph\rr \phi\ome$~\cite{Belle:2012qqr}. The phase space is expected to drop significantly near the threshold, leading to a decrease in the cross section. However, the contribution from the $0^+$ state does not seem to decrease near $2.0 \,\text{GeV}$, likely due to the contribution from $f_0(1710)$. Therefore, by fitting the partial wave data of $0^+$ contribution from Belle for $\ph\ph\rr \phi\ome$~\cite{Belle:2012qqr}, we can determine the value of $\Lam_3$, which represents the cutoff parameter applied in the dipole form factor for the vector mesons $\rho,\,\omega,\,\phi$ in the $\gamma V$ vertex. The corresponding Feynman diagram is shown in Fig.~\ref{fig:feynmanLam3}.

 Note that the two vector mesons $(V(k_1),V(k_2))$ represented at the $\gamma V$ vertex in Fig.~\ref{fig:feynmanLam3} include the following pairs: $(\rho^0(k_1),\rho^0(k_2))$, $(\phi(k_1),\phi(k_2))$, $(\omega(k_1),\omega(k_2))$, $(\phi(k_1),\omega(k_2))$, $(\omega(k_1),\phi(k_2))$. The corresponding scattering amplitude is presented in Appendix~\ref{sec:Appendix B}. The total width of $f_0(1710)$ is taken to be $0.150\,\text{GeV}$~\cite{PDG}. As for the mass, we adopt a range from $1.71\,\text{GeV}$~\cite{Wang:2021jub} to $1.78\,\text{GeV}$, which is the mass threshold of $K^*\bar K^*$.

Through data fitting, we obtained $\Lambda_3=0.63\,\text{GeV}$ when the mass of $f_0(1710)$ is $1.71\,\text{GeV}$, and $\Lambda_3=0.83\,\text{GeV}$ when the mass is $1.78\,\text{GeV}$. 
Note that here $\Lambda_3$ is especially sensitive to mass variations. This sensitivity arises because the coupling constant $g_{f_0K^*\bar{K}^*}$, calculated with Eq.~(\ref{eq33}), decreases significantly as the binding energy $\epsilon$ approaches 0.

\subsection{Double-Photon Luminosity Function}
If the cross section $\sigma_{\gamma\gamma}$ for a two-photon production reaction is known, and the integrated luminosity $\int Ldt$ for $e^+e^-$ scattering is provided, then we can obtain the event yield for the two-photon production as follows:
\begin{equation}Y_{\ph\ph}(W) = \sigma_{\gamma\gamma}(W) L_{\gamma\gamma}(W) \int Ldt \Delta W,
\end{equation}
where $W$ is the c.m.\ energy of the two incident photons. $L_{\dph}(W)$ is the double-photon luminosity function. $\Delta W$ is the bin width. $Y_{\dph}(W)$ represents the event yield for two-photon production in a $W$ range between $W$ and $W+\Delta W$.

Let $E_0$ be the initial energy of the electron in the $e^+e^-$ c.m.~frame. In the same frame, assume two photons with energies $E_1$ and $E_2$, and four-momentum squares $-Q^2$ and $-P^2$, are produced in the $e^+e^-$ scattering. For $Q^2\to0$ and $P^2\to0$, the luminosity function $L_{\gamma\gamma}$ is given as follows~\cite{Berger:1986ii}:
\begin{equation}
    L_{\gamma\gamma}(W) = \int f_{\gamma/e}(y,W,Q^2_{\text{max}}) f_{\gamma/e}(z,W,Q^2_{\text{max}})\frac{W}{2E_0^2} \frac{1}{2r} dr,
\end{equation}
where $y\approx E_{1}/E_0$, $z=E_{2}/E_0$, and $r=y/z$. 

The definition of $f_{\gamma/e}$ is given as follows\cite{Belle:2009xpa}:
\splittwo{f_{\gamma/e}(z,W,Q^2_{\text{max}}) = \frac{\alpha}{\pi z} \left\{ (1+(1-z)^2)\f12\right.}
{\left.\int_{ln\left(m_e^2z^2/(1-z)\right)}^{lnQ^2_{\text{max}}}F(e^\nu,W)d\nu -1+z \right\},}{f_ph function}
where $\alpha\approx\frac{1}{137}$ is the fine-structure constant. $\nu=\ln Q^2$ and $F(Q^2,W)=1/(1+Q^2/W^2)^2$\cite{Belle:2009xpa}. $Q^2_{\text{max}}=1\,\text{GeV}^2$ gives a maximum virtuality of the incident photons. Details on the derivation of the expression for $L_{\gamma\gamma}$ can be found in Appendix~\ref{sec:Appendix A}.

\section{Numerical results and discussion }\label{sec:Results}
\begin{figure}[htbp]
  \centering
  \begin{minipage}{\linewidth}
    \centering
    \includegraphics[width=0.99\linewidth]{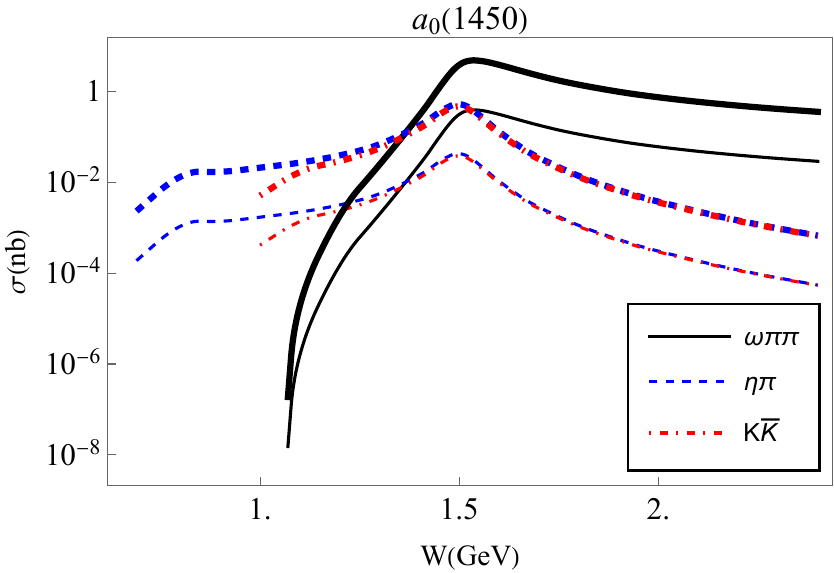} 
    \label{a0(1450)-cs}
    \includegraphics[width=0.99\linewidth]{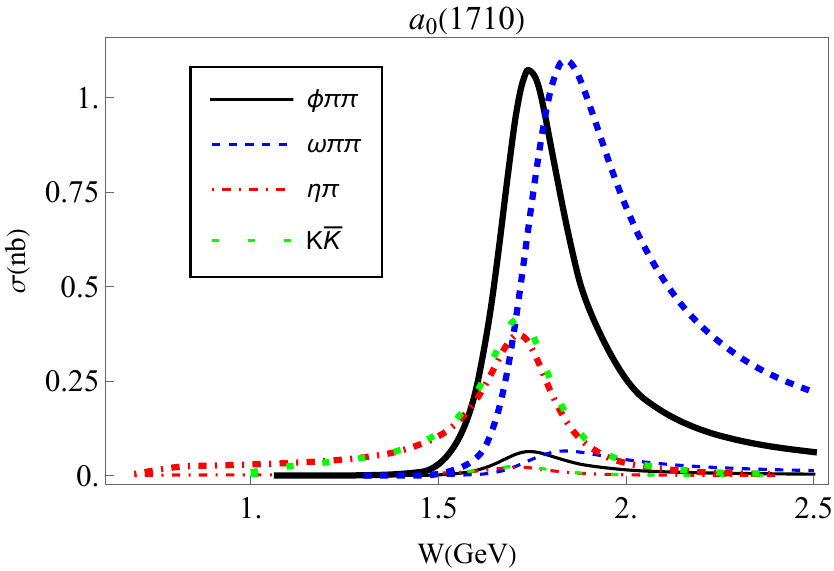}
    \label{fig:a0(1710)-cs}
    \includegraphics[width=0.99\linewidth]{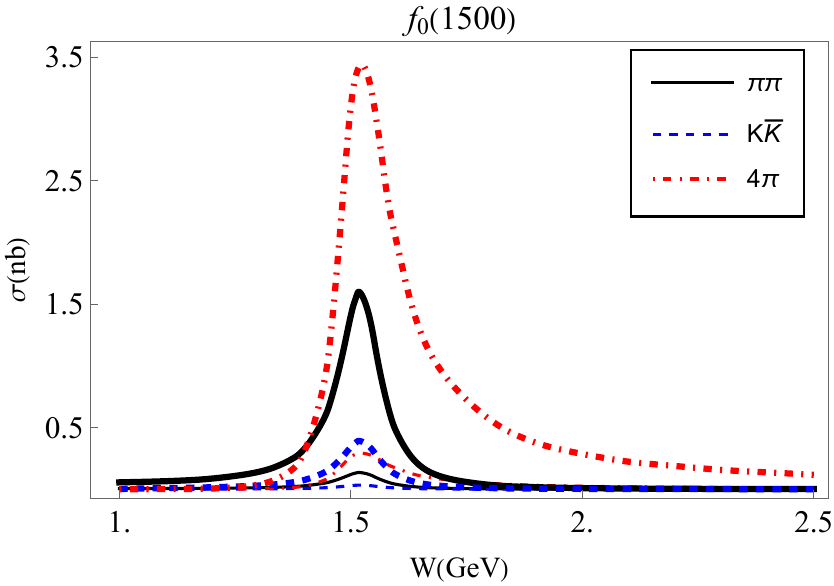} 
    \label{fig:f0(1500)-cs}
    \includegraphics[width=0.99\linewidth]{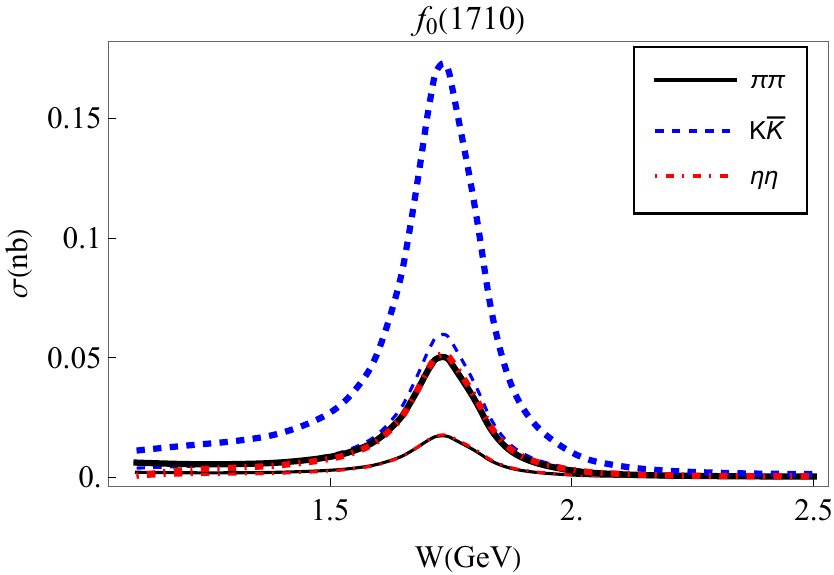}
    \label{fig:f0(1710)-cs}
  \end{minipage}
  \caption{Cross sections for two-photon production of $a_0(1450)$(Log scale), $a_0(1710)$, $f_0(1500)$, and $f_0(1710)$ vs. c.m. energy $W$. In each figure, the lower group of lines corresponds to $\Lambda_3=0.63\,\text{ GeV}$, and the upper group corresponds to $\Lambda_3=0.83\,\text{ GeV}$.}
  \label{fig:cs}
\end{figure}

Figure~\ref{fig:cs} shows the distribution of cross sections as a function of the c.m. energy $W$ of the incoming $\gamma\gamma$ system for the two-photon production of $a_0(1450)$, $a_0(1710)$, $f_0(1500)$, and $f_0(1710)$.

In this work, we use a fixed $e^+e^-$ c.m. energy of $10.58\,\text{GeV}$ for the double-photon luminosity function calculation. The total integrated luminosity is set to be $8000\,\text{fb}^{-1}$ for the prediction of $N_p$s at SuperKEKB, which is about 40 times the luminosity of Belle's $223\,\text{fb}^{-1}$ in their $\gamma\gamma\to\eta\pi^0$ analysis.~\cite{Belle:2009xpa}.

\begin{table*}[htb]
\centering
\label{Sum of tables}

\begin{minipage}{\textwidth}
\centering
\captionsetup{justification=centering}
\caption{Peak information related to $a_0(1450)$}
\label{a0(1450)Table}
\begin{tabular}{ccccc}
\toprule
   & $\omega\pi\pi$ & $\eta\pi$ & $K\bar K$ \\
\midrule
$E_p$ (Peak energy) [GeV] & $1.54$ & $1.50$ & $1.50$ \\
$\sigma_p$ (Peak cross section) [nb] & $0.39\text{--}4.77$ & $0.04\text{--}0.52$ & $0.04\text{--}0.46$\\
$N_p$ (Peak event yield) [$10^5/(10\,\text{MeV})$] & $1.53\text{--}18.8$ & $0.17\text{--}2.15$ & $0.16\text{--}1.92$ \\
\bottomrule
\end{tabular}
\end{minipage}
\vspace{3mm}

\begin{minipage}{\textwidth}
\centering
\captionsetup{justification=centering}
\caption{Peak information related to $a_0(1710)$}
\label{a0(1710)Table}
\begin{tabular}{ccccc}
\toprule
   & $\phi\pi\pi$ & $\omega\pi\pi$ & $\eta\pi$ & $K\bar K$ \\
\midrule
$E_p$ (Peak energy) [GeV] & $1.84$ & $1.74$ & $1.71$ & $1.71$ \\
$\sigma_p$ (Peak cross section) [nb] & $0.06\text{--}1.10$ & $0.06\text{--}1.07$ & $0.02\text{--}0.37$ & $0.02\text{--}0.41$ \\
$N_p$ (Peak event yield) [$10^5/(10\,\text{MeV})$] & $0.18\text{--}3.08$ & $0.20\text{--}3.35$ & $0.07\text{--}1.20$ & $0.08\text{--}1.34$ \\
\bottomrule
\end{tabular}
\end{minipage}
\vspace{3mm}

\begin{minipage}{\textwidth}
\centering
\captionsetup{justification=centering}
\caption{Peak information related to $f_0(1500)$}
\label{f0(1500)Table}
\begin{tabular}{ccccc}
\toprule
   & $\pi\pi$ & $K\bar K$ & $4\pi$\\
\midrule
$E_p$ (Peak energy) [GeV] & $1.52$ & $1.52$ & $1.52$ \\
$\sigma_p$ (Peak cross section) [nb] & $0.13$--$1.59$ & $0.03$--$0.39$ & $0.29$--$3.43$ \\
$N_p$ (Peak event yield) [$10^5/(10\,\text{MeV})$] & $0.55$--$6.44$ & $0.13$--$1.58$ & $1.18$--$13.9$ \\
\bottomrule
\end{tabular}
\end{minipage}
\vspace{3mm}

\begin{minipage}{\textwidth}
\centering
\captionsetup{justification=centering}
\caption{Peak information related to $f_0(1710)$}
\label{f0(1710)Table}
\begin{tabular}{ccccc}
\toprule
   & $\pi\pi$ & $K\bar K$ & $\eta\eta$\\
\midrule
$E_p$ (Peak energy) [GeV] & $1.74$ & $1.74$ & $1.74$ \\
$\sigma_p$ (Peak cross section) [nb] & $0.02$--$0.05$ & $0.06$--$0.17$ & $0.02$--$0.05$ \\
$N_p$ (Peak event yield) [$10^5/(10\,\text{MeV})$] & $0.05$--$0.16$ & $0.19$--$0.54$ & $0.06$--$0.16$ \\
\bottomrule
\end{tabular}
\end{minipage}
\vspace{3mm}
\end{table*}

Tables~\ref{a0(1450)Table} through~\ref{f0(1710)Table} present the peak information for $E_p$ (peak energy in $\text{GeV}$), $\sigma_p$ (peak cross section in $\text{nb}$), and $N_p$ (peak event yield in $10^5/(10\,\text{MeV})$) across reactions, with the values varying based on $\Lambda_3$. To avoid any confusion, it should be noted that the previously determined range for $\Lambda_3$, based on two sets of $f_0(1710)$ masses at $1.71\,\text{GeV}$ and $1.78\,\text{GeV}$, serves as a foundation for our calculations. However, the specific mass of $f_0(1710)$ employed in our cross section calculations is the average value in RPP, which is 1733 MeV, situated between these two mass values. Furthermore, given that the partial widths of $f_0(1710)$ used in our calculations are derived from the unitary coupled channel approach, we have also adopted a set of mass and width values for $f_0(1710)$ from~\cite{Wang:2021jub}. Consequently, the results are approximately 2.4 times those presented in Table~\ref{f0(1710)Table}, yet they remain within the same order of magnitude.

In Belle's analysis of $\gamma\gamma \to \eta\pi^0$, the efficiency $\eta$ for $W \in [1.0, 2.0]\,\text{GeV}$ generally lies between $0.01$ and $0.02$ ~\cite{Belle:2009xpa}. The event yield of $\gamma\gamma\to a_0(980)/a_0(Y)\to \eta\pi^0$ displays a notable peak at $W_0=0.9823\,\text{GeV}$ with event yield about $2000/(10\,\text{MeV})$, denoted as $N_0$\footnote{$N_0$ is not directly presented in the reference~\cite{Belle:2009xpa}, but we can estimate it with the following facts: Consider only the S-wave, the total cross section contributed by these two $a_0$ resonances is around $10\,\text{nb}$, while the total cross section including background is around $36\,\text{nb}$ at $W_0$, both integrated over $|\cos\theta^*|<0.8$. The total event yield for $|\cos\theta^*|<0.05$ is around $900/20\,\text{MeV}$ at $W_0$, and note that the S-wave contribution should be evenly distributed over $0<|\cos\theta^*|<0.8$. Therefore, we have $N_0=\left(\frac{1}{2} \frac{0.8}{0.05} \frac{10}{36} 900\right)/10\,\text{MeV} = 2000/10\,\text{MeV}$.
}.
Here $a_0(Y)$ represents an unidentified scalar resonance.

For comparison, the calculated $N_p$ for the reaction $\gamma\gamma \to a_0(1450) \to \omega\pi\pi$ at SuperKEKB ranges between $76.5\text{--}940N_0$. By setting $\eta=0.01$, the range of $N_p$ narrows to $0.8\text{--}9.4N_0$, still at or above the order of magnitude of $N_0$.
Hence, Belle-II is expected to clearly observe the resonance peak of $a_0(1450)$ through the reaction $\ph\ph \to \omega\pi\pi$ at SuperKEKB. Similarly, the calculated $N_p$ for the reaction $\gamma\gamma\to a_0(1450) \to \eta\pi$ at SuperKEKB ranges between $8.5\text{--}107.5N_0$. Setting $\eta=0.01$, $N_p$ narrows to $0.1\text{--}1.1N_0$, which approaches or reaches the order of magnitude of $N_0$. This suggests that observing the $a_0(1450)$ resonance peak through the reaction $\gamma\gamma \to \eta\pi^0$ at SuperKEKB is likely. Similarly, the expected $N_p$ of $8\text{--}96N_0$ for the reaction $\gamma\gamma\to K\bar K$ indicates a similar potential for observation.

From analogous analyses, the observation prospects for other resonances at SuperKEKB can be inferred. Table~\ref{a0(1710)Table} indicates that $a_0(1710)$ is likely observable via the reactions $\gamma\gamma\to\phi\pi\pi,\,\omega\pi\pi,\,\eta\pi,\,K\bar K$, with expected $N_p$ ranging between $9\text{--}154N_0$, $10\text{--}167.5N_0$, and $3.5\text{--}60N_0$, $4\text{--}67N_0$, respectively.
Table~\ref{f0(1500)Table} indicates that $f_0(1500)$ is observable via the reactions $\gamma\gamma \to \pi\pi,\,4\pi$, with expected $N_p$ ranging between $27.5\text{--}322N_0$, $59\text{--}695N_0$, and suggests a possible observation in the $K\bar{K}$ channel, estimated at $6.5\text{--}79N_0$.
Tables~\ref{f0(1710)Table} indicates a modest likelihood of observing $f_0(1710)$ in the reactions $\gamma\gamma \to \pi\pi,\, K\bar K,\,\eta\eta$, with expected $N_p$ ranging between $2.5\text{--}8N_0$, $9.5\text{--}27N_0$, $3\text{--}8N_0$.

Transitioning from Belle II's potential observations at SuperKEKB, we now consider the scenario at KEKB's original integrated luminosity. By setting the integrated luminosity to $223\,\text{fb}^{-1}$, the expected $N_p$ for $\dph \to a_0(1450)\to \eta\pi$ falls within $0.005\text{--}0.06N_0$ at $\eta=0.02$. For context, Belle has documented a resonance peak for $a_0(980)$, $a_0(Y)$ with an event yield of $N_0$, and the number of all candidate events at the peak is approximately $5N_0$. Thus, Belle's failure to observe an $a_0(1450)$ resonance peak in the reaction $\gamma\gamma \to \eta\pi^0$ aligns with our projection.

We proceed to compare our calculations with Belle's measurements on $f_0(1500)$ and $f_0(1710)$. In a detailed analysis of the reaction $\gamma\gamma \to \pi^0\pi^0$, Belle found no distinct resonance peak for the $f_0(1500)$ around $1.5\,\text{GeV}$. Nevertheless, the nominal fit suggests a possible contribution from $f_0(1370)$, $f_0(1500)$, or a mixed contribution from both, denoted as $f_0(Y)$. This fit yields a mass of $1469.7 \pm 4.7\,\text{MeV}/c^2$, a total decay width of $89.7^{+8.1}_{-6.6}\,\text{MeV}$, and $\Gamma_{\gamma\gamma}\mathcal{B}(f_0(Y) \to \pi^0\pi^0) = 11.2^{+5.0}_{-4.0}\,\text{eV}$.

Now we could estimate the cross section for the reaction $\gamma\gamma \to f_0(Y) \to \pi^0\pi^0$ with these parameters. Firstly, note that Belle adopts the relativistic Breit-Wigner resonance amplitude, denoted as $A_R(W)$, to fit the peak of a spin-J resonance $R$ with mass $m_R$, as described in the following equation~\cite{Belle:2008bmg}:
\begin{equation}
    A_R^J(W) = \sqrt{\frac{8\pi(2J+1)m_R}{W}}\frac{\sqrt{\Gamma_{\gamma\gamma}(W) \Gamma_{\pi^0\pi^0}(W)}}{m_R^2-W^2-im_R\Gamma_{\text{tot}}(W)}.
    \label{eq:ARW}
\end{equation}

Then we could obtain $\sigma_p=0.71\,\text{nb}$ at $W=1.47\,\text{GeV}$ with the following formula:
\begin{equation}
    \sigma(W) = \int d\Omega|A_{f_0(Y)}(W)|^2|Y^0_0|^2,
\label{cs formula}
\end{equation}
where $Y_0^0=\sqrt{\frac{1}{4\pi}}$, and $A_{f_0(Y)}(W)$ is given by Eq.~(\ref{eq:ARW}).

From our calculations, the $\sigma_p$ for the reaction $\gamma\gamma \to f_0(1500) \to \pi\pi$ is $0.13\text{--}1.59\,\text{nb}$ at $W=1.52\,\text{GeV}$. Therefore, the $\sigma_p$ for the reaction $\gamma\gamma \to f_0(1500) \to \pi^0\pi^0$ would be $0.04\text{--}0.53\,\text{nb}$, which is possibly comparable to the fit result $0.71\,\text{nb}$.

Belle has performed an analysis on the reaction $\gamma\gamma \to K\bar{K}$ to check the existence of the $f_0(1710)$ resonance within two fixed resonances $a_2(1320)$, $f_2(1270)$ and a floating resonance $f_2'(1525)$~\cite{Belle:2013eck}.
Six acceptable fits were identified for the $f_0(1710)$ resonance in this analysis.
For the first four solutions, Belle performed a fit for the c.m. energy region $1.2\,\text{GeV}\leq W \leq 2.0\,\text{GeV}$ by floating the mass, width, $\Gamma_{\gamma\gamma}\mathcal B(K\bar K)$ and the relative phase of both the $f_2'(1525)$ and $f_J(1710)$ ($J = 0$ or $J = 2$). This approach is referred to as Fit-All, with the four solutions denoted as fit-1, fit-2, fit-3, and fit-4.
Belle provided another two fits with different approach.
For these two solutions, Belle first obtained the $f_2'(1525)$ parameters by fitting in the range $1.15\,\text{GeV}\leq W \leq 1.65\,\text{GeV}$ and ignoring the contribution of the $f_J(1710)$. Then Belle included the contribution of the $f_0(1710)$ and fitted the region $1.2\,\text{GeV}\leq W \leq 2.0\,\text{GeV}$ by fixing the $f_2'(1525)$ parameters. This approach is referred to as Fit-Part, with the two solutions denoted as fit-H and fit-L.
We estimate the $\sigma_p$ for the reaction $\gamma\gamma \to f_0(1710) \to K\bar{K}$ based on these parameters, as detailed in Table~\ref{table:6 fits for f0(1710)}.

\begin{table}[htbp]
\centering
\caption{Peak cross sections for the reaction $\gamma\gamma\to f_0(1710)\to K\bar K$ in different fits. Mass is given in MeV/$c^2$, total width in MeV, and $\Gamma_{\gamma\gamma}\mathcal B(K\bar K)_{f_0}$ in eV.}
\label{table:6 fits for f0(1710)}
\begin{tabular}{c|cccc|cc}
\hline\hline
& \multicolumn{4}{c|}{Fit-All} & \multicolumn{2}{c}{Fit-Part} \\ 
\cline{1-7}
Sol. & fit-1 & fit-2 & fit-3 & fit-4 & fit-H & fit-L \\ 
\hline
Mass($f_0$) & 1781 & 1780 & 1783 & 1761 & $1750^{+5+29}_{-6-18}$ & $1749^{+5+31}_{-6-42}$ \\ 
$\Gamma_{\text{tot}}(f_0)$ & 99 & 110 & 96 & 119 & $138^{+12+96}_{-11-50}$ & $145^{+11+31}_{-10-54}$ \\ 
$\Gamma_{\gamma\gamma}\mathcal B(K\bar K)_{f_0}$ & 216 & 6.3 & 189 & 10.3 & $12^{+3+227}_{-2-8}$ & $21^{+6+38}_{-4-26}$ \\ 
\hline
$\sigma_p$ [nb] & 6.7 & 0.2 & 6.1 & 0.3 & $0.3^{+0.07+5.3}_{-0.05-0.2}$ & $0.5^{+0.14+0.8}_{-0.09-0.6}$ \\ 
\hline\hline     
\end{tabular}
\end{table}

From our calculations, the $\sigma_p$ for the reaction $\gamma\gamma \to f_0(1710) \to K\bar K$ is $0.17\,\text{nb}$ when $\Lambda_3=0.83\,\text{GeV}$, which falls within the same order of magnitude as the estimates from fit-2, fit-4, fit-H, and fit-L.

The two-photon decay width of $f_0(1710)$ could offer another valuable perspective.
Belle~\cite{Belle:2013eck} and CELLO~\cite{CELLO:1988xbx} both measured $\Gamma_{K\bar K}\Gamma_{\gamma\gamma}/\Gamma_{\text{tot}}$ in the reaction $\gamma\gamma\to K_S^0 K_S^0$. Utilizing $\Gamma_{K\bar K}/\Gamma_{\text{tot}}=0.38^{+0.09}_{-0.19}$~\cite{Longacre:1986fh}, we could obtain the corresponding $\Gamma_{\gamma\gamma}$, as listed in Table.~\ref{table:two-photon decay width of f0(1710)}. 
On the other hand, $f_0(1710)$ has been considered to be dynamically generated by the vector meson-vector meson interaction within the hidden gauge formalism, and calculations of the two-photon decay width of $f_0(1710)$ as well as its iso-vector partner are presented in~\cite{Branz:2009cv}. The results are listed alongside ours in Table.~\ref{table:two-photon decay width of f0(1710)} and Table.~\ref{table:two-photon decay width of a0(1710)}.

\begin{table}[htbp]
\centering
\caption{$\Gamma_{\gamma\gamma}$ for $f_0(1710)$ from experimental and theoretical studies}
\label{table:two-photon decay width of f0(1710)}
\begin{tabular}{l|l|c|c}
\hline\hline
\multicolumn{2}{c|}{Source} & $\Gamma_{\gamma\gamma}\mathcal B(K\bar K)\,[\text{eV}]$ & $\Gamma_{\gamma\gamma}\,[\text{eV}]$ \\ 
\hline
\multirow{2}{*}{Exp} & BELLE & $12^{+3+227}_{-2-8}$ & $32^{+8+597}_{-5-21}$ \\ 
                              & CELLO & $<110$ & $<289$ \\ 
\hline
\multirow{2}{*}{Th}  & Ref~\cite{Branz:2009cv} & $\backslash$ & $50$ \\ 
                              & Present work & $\backslash$ & $2\text{--}8$ \\ 
\hline\hline     
\end{tabular}
\end{table}

\begin{table}[htbp]
\centering
\caption{$\Gamma_{\gamma\gamma}$ for $a_0(1710)$ from theoretical studies}
\label{table:two-photon decay width of a0(1710)}
\begin{tabular}{l|c|c}
\hline\hline
\multicolumn{2}{c|}{Source} & $\Gamma_{\gamma\gamma}\,[\text{keV}]$ \\ 
\hline
\multirow{2}{*}{Th}  & Ref~\cite{Branz:2009cv}  & $1.61$ \\ 
                              & Present work  & $0.03\text{--}0.57$ \\ 
\hline\hline     
\end{tabular}
\end{table}

The error in Belle's measurement might be too large to be convincing, yet both our calculations and those presented in~\cite{Branz:2009cv} align with the CELLO data, despite our results being one to two orders of magnitude lower. This difference may in part be due to the form factor for the $\rho^0$, $\omega$, and $\phi$ mesons' coupling to photons introduced in our work, as shown in Eq.~(\ref{eq:dddd}) for the estimation of the total cross section. Such a form factor could notably reduce the two-photon decay width by a factor $r$, ranging between $0.006\text{--}0.08$.
This difference may in part be due to the form factor for the $\rho^0$, $\omega$, and $\phi$ mesons' coupling to photons introduced in our work, as shown in Eq.~(\ref{eq:dddd}), which participates in the calculation of the cross section and width. Such a form factor could notably reduce the two-photon decay width by a factor $r$, ranging between $0.006\text{--}0.08$.
\begin{equation}
    r=\left(\frac{(\Lambda_3)^4}{(\Lambda_3)^4+m_V^4}\right)^4,\label{eq:dddd}
\end{equation}
where we have assumed a general mass of vector mesons, $m_V=0.8\,\text{GeV}$.

The situation is similar for the $f_0(1500)$. Ref~\cite{Nagahiro:2008um} initially considered the $f_0(1370)$ to be dynamically generated by the $\rho\rho$ interaction, providing its two-photon decay width as $1.62\,\text{keV}$, without considering the form factor of the $\rho^0$, $\omega$, and $\phi$ mesons that couple to the photon, either. This identification of $f_0(1370)$ shifts to $f_0(1500)$ after including the coupled-channel effects of pseudoscalar mesons~\cite{Wang:2019niy}. In comparison, our results for the two-photon decay width of $f_0(1500)$ are in the range of $0.01\text{--}0.12\,\text{keV}$, which are also one to two orders of magnitude lower.

It should be noted that omitting the form factor would lead to even larger predictions for the cross section of the two-photon production process, thereby increasing the potential for observing the aforementioned $a_0$ and $f_0$ resonances.

\section{Summary}\label{sec:Summary}
We consider $a_0(1450)$ and $a_0(1710)$ as hadronic molecules and calculate the cross sections for the two-photon production reactions $\gamma\gamma \to a_0(1450) \to \omega\pi\pi,\, \eta\pi^0,\, K\bar K$, $\gamma\gamma \to a_0(1710) \to  \phi\pi\pi,\,\omega\pi\pi,\, \eta\pi^0,\, K\bar K$, $\gamma\gamma \to f_0(1500) \to \pi\pi,\, K\bar K,\, 4\pi$, and $\gamma\gamma \to f_0(1710) \to \pi\pi,\, K\bar K,\, \eta\eta$, along with the expected event yields at SuperKEKB.

Based on our calculations, the peak cross sections for the reactions $\gamma\gamma \to f_0(1500)\to \pi\pi$ and $\gamma\gamma \to f_0(1710) \to K\bar K$ might be comparable to the experimental values reported by Belle in their $\gamma\gamma \to \pi^0\pi^0$ and $\gamma\gamma \to K_S^0 K_S^0$ analysis~\cite{Belle:2008bmg,Belle:2013eck}. Moreover, our calculations indicate that the event yield for $\dph\rr a_0(1450)\rr \eta\pi^0$ is not sufficient to form an observable peak in KEKB, which is consistent with the observation by Belle~\cite{Belle:2009xpa}.

Based on our predictions, $a_0(1450)$ can be observed in the reaction $\gamma\gamma \to \omega\pi\pi$ and may be possibly observed in $\gamma\gamma \to \eta\pi^0$. $a_0(1710)$ could be possibly observed in $\gamma\gamma \to \phi\pi\pi,\,\omega\pi\pi,\,\eta\pi^0,\,K\bar K$ at SuperKEKB. Furthermore, observation of $a_0(1710)$ in reaction $\gamma\gamma \to \phi\pi\pi$ could serve as confirmation of its hidden strange quark content.

The reaction $\gamma\gamma \to \omega\pi\pi$ is expected to be the most suitable for observing $a_0(1450)$, while $\gamma\gamma \to \phi\pi\pi$ is expected to be most suitable for observing $a_0(1710)$.
If Belle-II successfully utilizes SuperKEKB to observe these reactions and measure the masses and decay widths of $a_0(1450)$ and $a_0(1710)$ more accurately, it will help us verify their hadronic molecule interpretations as $\rho\omega$ and $K^*\bar K^*$ molecular states, respectively.

The $f_0(1500)$ and $f_0(1710)$ were previously considered as glueball candidates~\cite{Giacosa:2005zt,Chanowitz:2005du,Chao:2007sk}. As iso-scalar glueballs, they would have no iso-vector partners. Therefore, confirming $a_0(1450)$ and $a_0(1710)$ as the iso-vector partners of $f_0(1500)$ and $f_0(1710)$ is crucial in determining their nature.

If the above hypotheses are successfully verified, it could be confirmed that the resonances $f_0(980)$, $f_0(1500)$, and $f_0(1710)$ in the $I=0$ sector are $K\bar K$, $\rho\rho$, and $K^*\bar K^*$ iso-scalar molecular states, respectively. Similarly, the resonances $a_0(980)$, $a_0(1450)$, and $a_0(1710)$ in the $I=1$ sector are $K\bar K$, $\rho\omega$, and $K^*\bar K^*$ iso-vector molecular states, respectively. These results are summarized in Table~\ref{Sum Table}.
\begin{table}[htb]
  \caption{Hadronic molecule table of $f_0$ and $a_0$ resonances}
  \label{Sum Table}
  \centering
  \[\begin{array}{|c|c|c|c|}
    \hline
 \text{} & \smash{\raisebox{-1.5pt}{$ K$}} \smash{\raisebox{-1.5pt}{$\bar K$}} & \rho\rho/\rho\omega & \smash{\raisebox{-1.5pt}{$ K^*$}}\smash{\raisebox{-1.5pt}{$\bar K^*$}}\\
    \hline
    I=0 & f_0(980) & f_0(1500) & f_0(1710) \\
    \hline
    I=1 & a_0(980) & a_0(1450) & a_0(1710) \\
    \hline
  \end{array}\]
\end{table}

\section*{Acknowledgments}

We thank useful discussions and valuable comments from Feng-Kun Guo and Hao-jie Jing. 
This work is supported by the NSFC and the Deutsche Forschungsgemeinschaft (DFG, German Research
Foundation) through the funds provided to the Sino-German Collaborative
Research Center TRR110 “Symmetries and the Emergence of Structure in QCD”
(NSFC Grant No. 12070131001, DFG Project-ID 196253076 - TRR 110), by the NSFC 
Grant No.11835015, No.12047503, No. 12175239, No. 12221005 and by the Chinese Academy of Sciences (CAS) under Grant No.XDB34030000,
and by the Chinese Academy of Sciences under Grant No. YSBR-101, and by the Xiaomi Foundation / Xiaomi Young Talents Program.


\appendix
\section{Supplement to Double-Photon Luminosity Function}\label{sec:Appendix A}
In this appendix we detail the derivation of the double-photon luminosity function.

$L_{\gamma\gamma}(W)$ is the probability distribution function of a two-photon production with a total energy of $W$ from a pair of beam particles, which refer to $e^+e^-$ in the context. Its physical significance can be seen from the following equation \cite{Uehara:1996bgt}:
\begin{equation}
\sigma_{e^+e^-} = \int\sigma_{\gamma\gamma}(W) L_{\gamma\gamma}(W)dW.
\label{17}
\end{equation}

Here, $\sigma_{e^+e^-}$ denotes the cross section for the reaction $e^+e^- \to e^+e^-X$, where $X$ can represent any final-state particles. 

To obtain the expression for $L_{\dph}$, we first need to derive the expression for $\sigma_{e^+e^-}$. Under the condition that the two incident photons are nearly on-shell ($Q^2\to 0$, $P^2\to 0$), Eq. (2.29) in \cite{Berger:1986ii} provides a simplified equation:
\begin{equation}
    \sigma_{e^+e^-} = \int f_{\gamma/e}(y,W,Q^2_{\text{max}}) f_{\gamma/e}(z,W,Q^2_{\text{max}})\sigma_{\gamma\gamma}(W)dydz.
\end{equation}

Define $s=yz=W^2/4E_0^2$ and recall that $r=y/z$. To simplify calculations, we can perform the integration variable transformation as follows: $dy\,dz \to J_{rs}\,dr\,ds$ and subsequently $ds \to J_{sW}\,dW$. This introduces two Jacobian factors: \begin{equation}
J_{rs}=\left|\frac{\partial(y,z)}{\partial(r,s)}\right|=\frac{1}{2r},
\end{equation}
and
\begin{equation}
J_{sW}=\left|\frac{\partial s}{\partial W}\right| = \frac{W}{2E_0^2}.
\end{equation}
Thus, we have:
\begin{equation}
    L_{\gamma\gamma}(W) = \int f_{\gamma/e}(y,W,Q^2_{\text{max}}) f_{\gamma/e}(z,W,Q^2_{\text{max}})\frac{W}{2E_0^2} \frac{1}{2r} dr,
\end{equation}
as have shown in \sref{sec:Formalism}.

The original formula of Eq.~(\ref{f_ph function}) is from Eq. (2.19) in \cite{Berger:1986ii}, and $\ln\frac{E(1-z)}{mz}\theta_{2,max}$ in Eq. (2.19) has been replaced with the integral \cite{Belle:2009xpa} :
 \begin{equation}\frac{1}{2}\int_{\ln\left(m_e^2z^2/(1-z)\right)}^{\ln Q^2_{\text{max}}} F(e^\nu,W)d\nu.
 \end{equation}

$F(Q^2,W)$ is introduced to describe the factorized relation between the cross sections for virtual photons and the real photons:
\begin{equation}
\sigma_{\gamma^*\gamma^*}(W)=F(Q_1^2,W)F(Q_2^2,W)\sigma_{\gamma\gamma}(W),
\end{equation}
where $\sigma_{\gamma^*\gamma^*}$ is the cross section for virtual photons, and $\sigma_{\gamma\gamma}$ is the cross section for real photons. 

\section{Supplement to Scattering Amplitudes}\label{sec:Appendix B}
In this appendix we present all the scattering amplitudes for each Feynman diagram given in \sref{sec:Formalism}.
For the reaction $\ph(k_1)\ph(k_2)\rr f_0(1500) \rr P(p_1)P(p_2)$ as shown in Fig.~\ref{fig:feynman1},
\begin{align}
\mathcal M_{1a} &= A_{1\mu}(m_1,k_1) A_{2\xi}(m_2,k_2) (F'_\rho)^2\nonumber \\ 
&\times (V_{\gamma(k_1)\rho})^{\mu\nu} (V_{\gamma(k_2)\rho})^{\xi\sigma}(G'_{\rho})_{\nu\alpha} (G'_{\rho})_{\sigma\beta} \nonumber \\ 
&\times (V_{f_0 \rho\rho})^{\alpha\beta} G_{f_0}(k_{1}+k_{2}) V_{f_0 P P}.
\end{align}

For the reaction $\ph(k_1)\ph(k_2)\rr f_0(1500) \rr \pi(p_1)\pi(p_2)\pi(p_3)\pi(p_4)$ as shown in Fig.~\ref{fig:feynman2},
\begin{align}
\mathcal M_{1b} &= A_{1\mu}(m_1,k_1) A_{2\xi}(m_2,k_2) (F'_\rho)^2\nonumber \\ 
&\times (V_{\gamma(k_1)\rho})^{\mu\nu} (V_{\gamma(k_2)\rho})^{\xi\sigma}(G'_{\rho})_{\nu\alpha} (G'_{\rho})_{\sigma\beta} \nonumber \\ 
&\times  (V_{f_0 \rho\rho})^{\alpha\beta}G_{f_0}(k_{1}+k_{2}) (V_{f_0 \rho\rho})^{\kappa\lambda}\nonumber
\\&\times(G_{\rho}(p_5))_{\kappa\delta}(G_{\rho}(p_6))_{\lambda\theta}(V_{\rho\pi\pi})^\delta(V_{\rho\pi\pi})^\theta.
\end{align}

For the reactions $\ph(k_1)\ph(k_2)\to f_0(1710) \to P(p_1)P(p_2)$ as shown in Fig.~\ref{fig:feynman3} and Fig.~\ref{fig:feynman4}, the scattering amplitudes $\mathcal{M}_{2a}$ and $\mathcal{M}_{2b}$ are nearly identical, with the exception that the $\gamma V$ vertex mesons $(\rho(k_1),\rho(k_2))$ in $\mathcal{M}_{2a}$ are replaced by $(\phi(k_1),\phi(k_2))$, $(\phi(k_1),\omega(k_2))$, $(\omega(k_1),\phi(k_2))$, and $(\omega(k_1),\omega(k_2))$ in $\mathcal{M}_{2b}$.

For the reaction $\ph(k_1)\ph(k_2) \to a_0(1450) \to P(p_1)P(p_2)$ as shown in Fig.~\ref{fig:feynman5},
\begin{align}
\mathcal M_{3a} &= A_{1\mu}(m_1,k_1) A_{2\xi}(m_2,k_2) F'_{V(k_1)}F'_{V(k_2)}\nonumber \\ 
&\times (V_{\gamma(k_1)V(k_1)})^{\mu\nu} (V_{\gamma(k_2)V(k_2)})^{\xi\sigma}(G'_{V(k_1)})_{\nu\alpha} (G'_{V(k_2)})_{\sigma\beta} \nonumber \\ 
&\times (V_{a_0 VV})^{\alpha\beta} G_{a_0}(k_{1}+k_{2}) V_{a_0 PP},
\end{align}
where $(V(k_1),\,V(k_2))$ includes $(\rho(k_1),\,\ome(k_2))$, $(\ome(k_1),\,\rho(k_2))$.

For the reaction $\ph(k_1)\ph(k_2) \to a_0(1450) \to \pi^+(p_1)\pi^-(p_2)\ome(p_3)$ as shown in Fig.~\ref{fig:feynman6},
\begin{align}
\mathcal M_{3b} &= A_{1\mu}(m_1,k_1) A_{2\xi}(m_2,k_2) F'_{V(k_1)}F'_{V(k_2)}\nonumber \\ 
&\times (V_{\gamma(k_1)V(k_1)})^{\mu\nu} (V_{\gamma(k_2)V(k_2)})^{\xi\sigma}(G'_{V(k_1)})_{\nu\alpha} (G'_{V(k_2)})_{\sigma\beta} \nonumber \\ 
&\times (V_{a_0 VV})^{\alpha\beta} G_{a_0}(k_{1}+k_{2}) (V_{a_0 \rho\omega})^{\kappa\lambda}\nonumber\\
&\times \epsilon_\kappa(m_3,p_3)(G_{\rho^0}(p_4))_{\lambda\delta}(V_{\rho^0\pi^+\pi^-})^\delta,
\end{align}
where $\epsilon_\kappa(m_3,p_3)$ is the polarization vector of the meson $\omega$.  

For the reaction $\ph(k_1)\ph(k_2) \to a_0(1710) \to P(p_1)P(p_2)$ as shown in Fig.~\ref{fig:feynman7},
\begin{align}
\mathcal M_{4a} &= A_{1\mu}(m_1,k_1) A_{2\xi}(m_2,k_2) F'_{V(k_1)}F'_{V(k_2)}\nonumber \\ 
&\times (V_{\gamma(k_1)V(k_1)})^{\mu\nu} (V_{\gamma(k_2)V(k_2)})^{\xi\sigma}(G'_{V(k_1)})_{\nu\alpha} (G'_{V(k_2)})_{\sigma\beta} \nonumber \\ 
&\times (V_{a_0 VV})^{\alpha\beta} G_{a_0}(k_{1}+k_{2}) V_{a_0 PP},
\end{align}
where $(V(k_1),\,V(k_2))$ includes $(\rho(k_1),\,\rho(k_2))$, $(\rho(k_1),\,\phi(k_2))$, $(\phi(k_1),\,\rho(k_2))$, $(\rho(k_1),\,\ome(k_2))$, $(\ome(k_1),\,\rho(k_2))$.

For the reaction $\ph(k_1)\ph(k_2) \to a_0(1710) \to \pi^+(p_1)\pi^-(p_2)\ome(p_3)$ as shown in Fig.~\ref{fig:feynman8},
\begin{align}
\mathcal M_{4b} &= A_{1\mu}(m_1,k_1) A_{2\xi}(m_2,k_2) F'_{V(k_1)}F'_{V(k_2)}\nonumber \\ 
&\times (V_{\gamma(k_1)V(k_1)})^{\mu\nu} (V_{\gamma(k_2)V(k_2)})^{\xi\sigma}(G'_{V(k_1)})_{\nu\alpha} (G'_{V(k_2)})_{\sigma\beta} \nonumber \\ 
&\times (V_{a_0 VV})^{\alpha\beta} G_{a_0}(k_{1}+k_{2}) (V_{a_0 \rho\omega})^{\kappa\lambda}\nonumber\\
&\times \epsilon_\kappa(m_3,p_3)(G_{\rho^0}(p_4))_{\lambda\delta}(V_{\rho^0\pi^+\pi^-})^\delta,
\end{align}
where $\epsilon_\kappa(m_3,p_3)$ is the polarization vector of the mesons $\phi$ or $\omega$.  

For the reaction $\ph(k_1)\ph(k_2)\to f_0(1710) \to \phi(p_1)\ome(p_2)$ as shown in Fig.~\ref{fig:feynmanLam3},
\begin{align}
\mathcal M_{5} &= A_{1\mu}(m_1,k_1) A_{2\xi}(m_2,k_2)F'_{V(k_1)}F'_{V(k_2)} \nonumber \\ 
&\times (V_{\gamma(k_1)V(k_1)})^{\mu\nu}(G'_{V(k_1)})_{\nu\alpha} \nonumber \\
&\times (V_{\gamma(k_2)V(k_2)})^{\xi\sigma} (G'_{V(k_2)})_{\sigma\kappa} \nonumber \\ 
&\times (V_{f_0 K^* \bar K^*})^{\delta\eta} G_{f_0}(k_{1}+k_{2}) (V_{f_0 K^* \bar K^*})^{\delta'\eta'}\nonumber\\
&\times \epsilon_{\alpha'}(m_1',p_1)\epsilon_{\kappa'}(m_2',p_2) G(k_1,k_2,k_5) \nonumber\\
&\times\int\frac{id^4 k_5}{(2\pi)^4} (-\tilde g_{\beta\delta}(k_{3}))(- \tilde g_{\lambda\eta}(k_{4}))(-\tilde g_{\gamma\theta}(k_{5}))\nonumber\\
&\times \left\{ (V_{V(k_1) K^{*+}K^{*-}} )^{\alpha\beta\gamma} (V_{V(k_2) K^{*-}K^{*+}})^{\kappa\lambda\theta}\right.\nonumber\\
&+ (V_{V(k_1) K^{*-}K^{*+}} )^{\alpha\beta\gamma} (V_{V(k_2) K^{*+}K^{*-}})^{\kappa\lambda\theta}\nonumber\\
&+ (V_{V(k_1) K^{*0}\bar K^{*0}})^{\alpha\beta\gamma} (V_{V(k_2)\bar K^{*0}K^{*0}})^{\kappa\lambda\theta}\nonumber\\
&+ \left.(V_{V(k_1)\bar K^{*0} K^{*0}})^{\alpha\beta\gamma} (V_{V(k_2) K^{*0}\bar K^{*0}})^{\kappa\lambda\theta}\right\} \nonumber\\
&\times\int\frac{id^4 p_5}{(2\pi)^4} (-\tilde g_{\beta'\delta'}(p_{3}))(- \tilde g_{\lambda'\eta'}(p_{4}))(-\tilde g_{\gamma'\theta'}(p_{5})) \nonumber\\
&\times G(p_1,p_2,p_5)\left\{ (V_{\phi K^{*+}K^{*-}} )^{\alpha'\beta'\gamma'} (V_{\ome K^{*-}K^{*+}})^{\kappa'\lambda'\theta'}\right.\nonumber\\
&+ (V_{\phi K^{*-}K^{*+}} )^{\alpha'\beta'\gamma'} (V_{\ome K^{*+}K^{*-}})^{\kappa'\lambda'\theta'}\nonumber\\
&+ V_{\phi K^{*0}\bar K^{*0}})^{\alpha'\beta'\gamma'} (V_{\ome\bar K^{*0}K^{*0}})^{\kappa'\lambda'\theta'}\nonumber\\
&+ \left.(V_{\phi\bar K^{*0} K^{*0}})^{\alpha'\beta'\gamma'} (V_{\ome K^{*0}\bar K^{*0}})^{\kappa'\lambda'\theta'}\right\} 
\end{align}
where $\epsilon_{\alpha'}(m_1',p_1)$ and $\epsilon_{\kappa'}(m_2',p_2)$ are the polarization vectors of the mesons $\phi$ and $\omega$.
\end{document}